# Two-dimensional Hydrogen-like Atom in a Constant Magnetic Field

M.G. Naber
mnaber@monroeccc.edu
Department of Science and Mathematics
Monroe County Community College
1555 S. Raisinville Rd
Monroe, Michigan, 48161-9746

The two-dimensional hydrogen-like atom in a constant magnetic field is considered. It is found that this is actually two separate problems. One for which the magnetic field causes an effective attraction between the nucleus and the electron and one for which it causes an effective repulsion. Each of the two problems has three separate cases depending on the sign of a shifted energy eigenvalue. For two of the six possibilities (shifted energy eigenvalue that is negative) it is shown that the first four solutions can be obtained exactly. For another two of the six possibilities (shifted energy eigenvalue that is positive) it is shown that the first eight solutions can be obtained exactly. For higher order states the energy eigenvalue is the root of a fifth or higher order polynomial, hence, the eigenvalue must be obtained numerically. Once the energy eigenvalue is known the solution to the radial wave equation is also known. Exact solutions for the radial wave equation, for the remaining two possibilities (shifted energy eigenvalue that is zero), are given to any desired order by means of a recursion relation.

## I. INTRODUCTION

The two-dimensional hydrogen-like atom in a constant magnetic field is a longstanding theoretical problem, for which only a few exact solutions are known[1,2], and has a long-standing history of perturbative study[3-9]. In recent years it has become an important problem for applications: single layer materials, quantum wells, etc. (see the references in Refs. 2, 7, and 10). Solutions of the two-dimensional problem are useful in sheading light on the problem of the three-dimensional hydrogen-like atom in a magnetic field as this is a non-separable and non-integrable problem[11-13]. In this paper it will be shown that the Schrödinger equation for the two-dimensional hydrogen-like atom in a constant magnetic field can be solved analytically by separating the problem into six different classes.

This study will take the point of view that a hydrogen-like atom exists in a two-dimensional plane and a magnetic field, perpendicular to this plane, is turned on slowly. Once the magnetic field attains the desired value it shall be taken to be constant. In this problem the energy of the electron comes from two sources: the electric field of the nucleus and the applied magnetic field. In this paper only states where the electron is bound to the nucleus are being sought. Electric field only bound states have negative energy. Magnetic field only bound states have positive energy. Both fields being on could have bound states with either positive or negative energy or even with no net energy. Additionally, there is an issue with a lack of symmetry for the full problem in that positive and negative $m$ values cause forces in opposite directions (the magnetic field is chosen to be along the positive $z$ axis and $m$ is the azimuthal quantum number). For positive $m$ values the magnetic and electric forces are parallel and for negative $m$ values the magnetic and electric forces are anti-parallel. Consideration of this lack of symmetry causes a deeper understanding of the magnetic field only system in which only positive or only negative $m$ values can occur depending on the orientation of the magnetic field.

In this paper exact solutions are found by partitioning the problem into separate cases according to the sign of a shifted energy eigenvalue as this can be positive, negative, or zero. This will allow for more asymptotic behavior to be removed from the radial differential equation. It should be noted that the magnetic field is arbitrary in strength but not in orientation and no approximations are used to find solutions.

In Sec. II a brief review of the two special cases for the problem are discussed with a careful examination of the magnetic field only case. In Sec. III the Schrödinger equation is derived and then partitioned into the six different possible cases representing the problem at hand. In Sec. IV the radial equations are solved for the first few states (or the single state as is the situation for cases 2 and 5). In the conclusion, with a summary of the possible types of energy eigenvalues and a discussion of solvability with regard to the Abel-Ruffini theorem, a question originally posed by Robnik and Romanovski[8] is answered.

## II. LIMITING CASES

The solution to the full problem must reduce to the known solutions for a hydrogen-like atom in two-dimensions and to an electron in a constant magnetic field. Both of these problems have long standing solutions and the latter problem is typically an example in most introductory books on quantum mechanics (cf. Ref. 14).

The Schrödinger equation for hydrogen-like atoms in two-dimensions is given by

$$\left(-\frac{\hbar^2}{2\mu}\nabla^2 - \frac{Ze^2}{4\pi\varepsilon_0}\frac{1}{r}\right)\psi = E\psi, \tag{1}$$

where $\mu = \frac{m_e M}{m_e + M}$ is the reduced mass, $m_e$ is the electron mass, $M$ is the mass of the nucleus, and $Ze$ is the positive charge on the nucleus. Throughout this paper notation and units will be the same as that used in Gasiorowicz[14]. The Coulomb potential is attractive and negative hence for bound states the energy, $E$, must be negative as well. In polar coordinates the Schrödinger equation becomes

$$\left(\partial_r^2 + \frac{1}{r}\partial_r + \frac{1}{r^2}\partial_\theta^2 + \frac{2\mu Ze^2}{4\pi\varepsilon_0\hbar^2}\frac{1}{r}\right)\psi - \frac{2\mu|E|}{\hbar^2}\psi = 0. \tag{2}$$

This equation is separable. Let $\psi = R(r)e^{im\theta}$, then,

$$\partial_r^2 R + \frac{1}{r}\partial_r R - \frac{m^2}{r^2}R + \frac{2\mu Z e^2}{4\pi\varepsilon_0 \hbar^2}\frac{1}{r}R - \frac{2\mu|E|}{\hbar^2}R = 0, \tag{3}$$

where $m$ is an integer (the azimuthal quantum number) characterizing the angular momentum about the "unobserved" $z$ axis. The third term can be thought of as a repulsive centrifugal barrier (cf. page 140 of Ref. 14). This equation is solved in any number of sources. A good reference for this is Englefield[15] where a comparison with the harmonic oscillator and group theoretic discussion is given. The energy eigenvalues are given by

$$E_C = -\frac{2\mu}{N_c^2}\left(\frac{Ze^2}{4\pi\varepsilon_0\hbar}\right)^2, \tag{4}$$

where $N_c = 2n_r + 2|m| + 1$, and $n_r$ is the radial quantum number and the subscript $c$ indicates Coulomb ($N_c$ is the principle quantum number). Notice that $N_c$ is always an odd number, the degeneracy of each energy level is equal to $N_c$, and, the energy is invariant under $m \to -m$. All solutions are of the form[16]

$$R_{n,m} = r^{|m|}e^{-\alpha r}L_{n-|m|}^{2|m|}\left(2\sqrt{|E|}r\right), \tag{5}$$

where, $\alpha = \sqrt{2\mu|E|}/\hbar$, and, $L_{n-|m|}^{2|m|}$ are the associated Laguerre polynomials.

The azimuthal quantum number, $m$, plays a very important role in determining the energy for the case of the combined coulomb and magnetic field, hence, some extra time will be spent considering the energy expression for the case of the magnetic field alone. The Schrödinger equation for an electron in a constant magnetic field can be derived as follows. Take the magnetic field to be along the $z$ axis (i.e. perpendicular to the plane of motion), that is, a positive magnetic field in the direction of the unobserved $z$ axis

$$\vec{B} = (0,0,B), \tag{6}$$

where $B > 0$. The Schrödinger equation is then ($\mu$ is now the electron mass)

$$\left(-\frac{\hbar^2}{2\mu}\nabla^2 - \frac{ie\hbar}{\mu}\vec{A}\cdot\vec{\nabla} + \frac{e^2}{2\mu}\vec{A}\cdot\vec{A}\right)\psi = E\psi. \tag{7}$$

One choice for the vector potential is

$$\vec{A} = -\frac{\vec{r}}{2}\times\vec{B}. \tag{8}$$

In polar coordinates Eq. (7) is separable, let $\psi = R(r)e^{im\theta}$,

$$\left(\partial_r^2 + \frac{1}{r}\partial_r - \frac{m^2}{r^2} - \frac{emB}{\hbar} - \left(\frac{eB}{2\hbar}\right)^2 r^2\right) R = -\frac{2\mu E}{\hbar^2} R. \tag{9}$$

Before proceeding it is useful to consider $m$ and its possible signs. Consider an electron moving in an empty two-dimensional plane at a constant velocity (the reader is asked to think semi-classically for a moment). As this plane is empty (there are no forces on the electron) there is no preferred location for the origin of any coordinate system. Let us suppose that the previously described magnetic field is slowly turned on from zero until it reaches a value of $B$. The magnetic force on the (negatively charged electron) causes the electron to begin to move on a circle in a counter clockwise fashion (viewed from above). i.e. $m$ must be positive in the Schrödinger equation given above (Eq. (9)). The center of the circle the electron is traveling on gives a preferred location for the origin of the coordinate system being used. Hence, when solving Eq. (9), the azimuthal quantum number can only be positive. If the magnetic field were turned on in the opposite direction then the electron would be going around a circle in a clockwise fashion and only negative $m$ values would be possible. Notice that the product of the allowed azimuthal quantum number and the given magnetic field will always be the same sign, hence, for either orientation of the magnetic field the Schrödinger equation is the same. The energy for both types of magnetic fields and azimuthal quantum numbers is

$$E_B = \frac{e|B|\hbar}{2\mu} N_B, \tag{10}$$

$$N_B = 2n_r + 2|m| + 1. \tag{11}$$

$n_r$ is the radial quantum number and the subscript $B$ indicates the magnetic field. Notice that $N_B$ is always an odd number. The degeneracy for positive $m$ values is $(N_B + 1)/2$ and likewise for the case of a reversed magnetic field and negative $m$ values. When the Coulomb field is combined with the magnetic field both positive and negative $m$ values can occur, for either orientation of the magnetic field, and, each will have a different effect on the combined system, one being attraction toward the nucleus and the other being repulsion from the nucleus. All solutions for the magnetic field only case are of the form (cf. page 252 of Ref. 14)

$$R_{n_r,m} = r^{|m|} e^{-\frac{\beta r^2}{4}} L_{n_r}^{|m|}(\beta r^2), \tag{12}$$

where, $\beta = e|B|/\hbar$ and $L_{n_r}^{|m|}$ are the associated Laguerre polynomials. The radial wave function is the product of three terms; a centrifugal term $r^{|m|}$, a term due to the magnetic field $e^{-\beta r^2/4}$, and a polynomial that depends on both $m$ and $\beta$.

### III. THE COULOMB AND MAGNETIC FIELD

In this problem start with a hydrogen-like atom, in any state, that is, with either a positive or negative $m$ value, and turn the magnetic field on slowly. Without loss of generality the magnetic field is taken to be along the positive $z$ axis. In this configuration positive $m$ values will represent states where the electric and magnetic attractions are parallel (i.e. the electron will have an increased attraction toward the nucleus due to the magnetic field). Negative $m$ values will represent states where the electric and magnetic attractions are anti-parallel (i.e. the magnetic field will cause a repulsion for the electron from the nucleus). Hence there will be an upper limit on $|m|$ (for negative $m$ values), that depends on $B$ and $Z$, after which the electron will become unbound from the nucleus. It is not meant that the electron will be in an unbound state, but, rather, a new bound state that might not contain the nucleus inside the "orbit". See Robnik and Romanovski[9] for a discussion of this topic in two dimensions and Friedrich and Wintgen[11] for a similar discussion in three dimensions. An interesting time dependent problem to consider would be to allow the magnetic field to be time dependent such that, $B(t = 0)$ is below the threshold value and $B(t = 1)$ be above the threshold value and numerically observe how the wave function transitions from the initial stable state to a later stable state.

The Schrödinger equation with both the Coulomb potential and the magnetic field has the form, using Eq. (8) to account for the magnetic field,

$$\left(-\frac{\hbar^2}{2\mu}\nabla^2 + \frac{e\hbar m}{2\mu}B + \frac{e^2 B^2}{8\mu}r^2 - \frac{Ze^2}{4\pi\varepsilon_0}\frac{1}{r}\right)\psi = E\psi. \tag{13}$$

This equation is separable in polar coordinates, let $\psi = R(r)e^{im\theta}$,

$$\left(\partial_r^2 + \frac{1}{r}\partial_r - \frac{m^2}{r^2} - \frac{emB}{\hbar} - \left(\frac{eB}{2\hbar}\right)^2 r^2 + \frac{2\mu Ze^2}{4\pi\varepsilon_0 \hbar^2}\frac{1}{r}\right)R = -\frac{2\mu E}{\hbar^2}R. \tag{14}$$

To keep the computations as clean as possible make the following substitutions. The first is a rescaling of the magnetic field, the second captures all the constants associated with the coulomb interaction, and the last is a rescaled energy. The dimensions of $\beta^{-1}$ and $\varepsilon^{-1}$ are area and the dimension of $\alpha^{-1}$ is length,

$$\beta = \frac{eB}{\hbar} > 0, \qquad \alpha = \frac{2\mu Ze^2}{4\pi\varepsilon_0 \hbar^2} > 0, \qquad \varepsilon = \frac{2\mu E}{\hbar^2}. \tag{15}$$

Note that $\varepsilon$ can be positive, negative, or zero. This introduces some mathematical difficulties that were not present in the two previous special cases (Eqs. (1) and (7)) The previous energy expressions (Eqs. (4) and (10)) now take the following form

$$\varepsilon_C = -\frac{\alpha^2}{N_C^2}, \qquad \varepsilon_B = \beta N_B. \tag{16}$$

Eq. (14) is now

$$\left(\partial_r^2 + \frac{1}{r}\partial_r - \frac{m^2}{r^2} - m\beta - \frac{\beta^2}{4}r^2 + \frac{\alpha}{r}\right)R = -\varepsilon R. \tag{17}$$

This equation can be cleaned up further using a dimensionless coordinate. Let $x = \sqrt{\beta}r$, $a = \alpha/\sqrt{\beta}$, and $\mathcal{E} = \varepsilon/\beta$ (i.e. distance is now measured in terms of Landau radii[8]). Then the equation becomes

$$\left(\partial_x^2 + \frac{1}{x}\partial_x - \frac{m^2}{x^2} - m - \frac{x^2}{4} + \frac{a}{x}\right)R = -\mathcal{E}R. \tag{18}$$

All terms in the above equation are dimensionless, including $a$ and $\mathcal{E}$ (note that a different parameter is used in this paper than that of Ref. 12). Properly speaking $\mathcal{E}$ should be written as $\mathcal{E}_{n,m}$ as the energy will depend on a radial and the azimuthal quantum number. Primes are now used to denote derivatives with respect to $x$,

$$R'' + \frac{1}{x}R' + \left(\mathcal{E} - m - \frac{m^2}{x^2} - \frac{x^2}{4} + \frac{a}{x}\right)R = 0. \tag{19}$$

To start the solution process, remove the short distance behavior (the centrifugal term). Let $R = x^{|m|}F(x)$, then,

$$F'' + \frac{(2|m| + 1)}{x}F' + \left(\mathcal{E} - m - \frac{x^2}{4} + \frac{a}{x}\right)F = 0. \tag{20}$$

Some far field behavior can be removed as well. Let $F = e^{-\frac{x^2}{4}}G(x)$, then,

$$G'' + \left(\frac{(2|m| + 1)}{x} - x\right)G' + \left(\mathcal{E} - m - |m| - 1 + \frac{a}{x}\right)G = 0. \tag{21}$$

Recall that $m$ can be either positive or negative and note that Eq. (21) is not invariant under $m \to -m$. These two choices give, respectively,

$$G'' + \left(\frac{2m + 1}{x} - x\right)G' + \left(\mathcal{E} - 2m - 1 + \frac{a}{x}\right)G = 0, \tag{22}$$

$$G'' + \left(\frac{2|m| + 1}{x} - x\right)G' + \left(\mathcal{E} - 1 + \frac{a}{x}\right)G = 0. \tag{23}$$

The reason for writing Eq. (21) as two different equations is that positive and negative $m$ values represent two different physical systems. Physically, Eq. (22) is for the magnetic field effectively causing an attraction between the electron and the nucleus. Eq. (23) is for the magnetic field effectively causing a repulsion between the electron and the nucleus. For Eq. (22) the magnetic field and/or the $m$ value can be as large as we wish but for Eq. (23) there is an upper limit for the effective magnetic force after which the electron will no longer be bound to the nucleus[8,9] (to be clear, the upcoming recursion relations that will generate solutions will allow for solutions beyond the threshold limit[8,9] but these solutions must be regarded as unphysical). To reiterate, this is not to say the electron will be in an unbound state, but, rather, a bound state whose "orbit" might not contain the nucleus. Another reason is mathematical. For linear second order differential equations, a term with a negative constant multiplying the unknown dependent variable generates an exponential function as part of the solution, hence when $\mathcal{E} - 2m - 1$ or $\mathcal{E} - 1$ is negative some of the asymptotic behavior can be accounted for. Notice that both Eq. (22) and Eq. (23) can be made to be formally self-adjoint and as such for a given $m$ value the eigenfunctions will be mutually orthogonal with respect to the appropriate weighting function (cf. page 63 of Ref. 17). Each equation represents a different physical system and as such the eigenfunctions for Eq. (22) and Eq. (23), while they might look similar, are part of different sets of orthogonal functions.

The eigenvalue that is sought is in $\mathcal{E}$, but there is another constant term in each equation that effectively shifts $\mathcal{E}$. This is, $\mathcal{E} - (2m + 1)$ for positive $m$ values and $\mathcal{E} - 1$ for negative $m$ values. It can now be seen that each equation will break into three separate cases (each with different solution types). For positive $m$ values (magnetic interaction is attractive)

$$1) \quad \mathcal{E} - (2m + 1) < 0, \tag{24}$$

$$2) \quad \mathcal{E} - (2m + 1) = 0, \tag{25}$$

$$3) \quad \mathcal{E} - (2m + 1) > 0, \tag{26}$$

and for negative $m$ values (magnetic interaction is repulsive)

$$4) \quad \mathcal{E} - 1 < 0, \tag{27}$$

$$5) \quad \mathcal{E} - 1 = 0, \tag{28}$$

$$6) \quad \mathcal{E} - 1 > 0. \tag{29}$$

## IV. SOLUTIONS

Given the different classes of solutions, some notational distinctions will be needed. Let $_aR_{n,m}$ and $_a\mathcal{E}_{n,m}$ represent the radial component of the wave function (not normalized) and the energy eigenvalue respectively. The leading index, $a$, will identify the solution case (1 through 6 corresponding to Eqs. (24) through (29)) and the subscripts, $n, m$, will identify a radial and azimuthal quantum number, respectively. As the solution method is essentially the same for each possible sign of the shifted energy eigenvalue the solutions will be presented in pairs. The reader is cautioned to remember that once $B$ and $Z$ are chosen there is an upper limit to $|m|$ for the $m < 0$ cases (for physical solutions).

Cases (1) and (4): $\mathcal{E} - (m + |m| + 1) < 0$

While cases (1) and (4) are physically different they can be solved by the same method. Let $\gamma^2 = |\mathcal{E} - m - |m| - 1|$ and $M = 2|m| + 1$. For the previously mentioned reasons $\gamma$ should be written as $\gamma_{n,m}$ and will be written as such for specific solutions. Eqs. (22) and (23) now become

$$G'' + \left(\frac{M}{x} - x\right)G' + \left(-\gamma^2 + \frac{a}{x}\right)G = 0. \tag{30}$$

Due to the sign of the $\gamma^2$ term one more piece of asymptotic behavior can now be removed. Let $G = e^{-\gamma x} H(x)$, then,

$$H'' + \left(\frac{M}{x} - x - 2\gamma\right)H' + \left(\frac{a - \gamma M}{x} + x\gamma\right)H = 0. \tag{31}$$

All obvious asymptotic behavior has been removed. A power series solution can now be sought. Let

$$H = \sum_{n=0}^{\infty} a_n x^n. \tag{32}$$

This generates three equations

$$a_1 = -\frac{(a - \gamma M)}{M} a_0, \tag{33}$$

$$a_2 = -\frac{(a - \gamma(M + 2))}{2(M + 1)} a_1 = \frac{(a - \gamma(M + 2))(a - \gamma M)}{2(M + 1)M} a_0, \tag{34}$$

$$a_n = -\frac{a_{n-1}\bigl(a - \gamma(M + 2(n - 1))\bigr) - a_{n-2}(n - 2) + \gamma a_{n-3}}{n(n - 1 + M)}. \tag{35}$$

Usually for eigenvalue/eigenfunction problems like those encountered in quantum mechanics (i.e. Sturm-Liouville problems on a semi-infinite or infinite domain) finite order polynomials are needed so that the overall wave functions are square integrable. Typically, this is because there is an overall factor of $e^{-Ax}$ or $e^{-Ax^2}$ (for some $A > 0$) as part of the radial wave function. The recursion relations for such problems have the freedom to choose a specific coefficient of the unknown power series to be zero. Once this choice is made the eigenvalue is computed and the corresponding eigenfunction (a polynomial of finite order) determined. The recursion relations (Eqs. (33)-(35)) for this current problem also have the same freedom to choose a specific coefficient of the power series to be zero, however, there is an overall factor of $e^{-\gamma x}e^{-\frac{x^2}{4}}$, hence, finite order polynomials are not necessarily needed to have a square integrable wave function. All that is needed are functions that grow no faster than $e^{Ax^2}$, where $A < 1/4$ is some positive constant. Indeed, this set of recursion relations (Eqs. (33)-(35)), do not necessarily generate finite order polynomials. Solutions to the differential equation can be found by the usual procedure of setting individual coefficients of the power series to zero to obtain the eigenvalue and then the subsequent eigenfunction.

The lowest order coefficient that can be set to zero is $a_1$ (which also forces $a_2$ to be zero but not $a_3$ and necessarily beyond). This gives,

$$a - \gamma M = 0, \tag{36}$$

$$\gamma_{1,m} = \frac{a}{M}, \tag{37}$$

$$_1\varepsilon_{1,m} = \beta(2m + 1) - \frac{\alpha^2}{(2m+1)^2}. \tag{38}$$

This has the correct limits (turning off the Coulomb field or turning off the magnetic field) with the radial quantum number being zero. The corresponding energy for case (4) is then,

$$_4\varepsilon_{1,m} = \beta - \frac{\alpha^2}{(2|m|+1)^2}. \tag{39}$$

Eq. (39) has the correct limit when turning off the magnetic field but physically would make no sense in the limit of turning off the electric field as negative $m$ values are not possible for the chosen orientation of the magnetic field (see the discussion in Sec. II concerning the magnetic field only case). The $H$ function for this choice is then (the subscript will reference the lowest order coefficient of the power series chosen to be zero)

$$H_1 = a_0 \left(1 - \frac{\gamma}{3(2+M)} x^3 + \frac{(a - \gamma(M+6))\gamma}{4(3+M)3(2+M)} x^4 - \cdots \right). \tag{40}$$

Where $\gamma$ is given by Eq. (37) and $a_5$ and beyond are given by Eq. (35) with the appropriate $\gamma$ and $M$ values being used. The associated radial component of the wave function is then, not normalized,

$$_{1 \text{ or } 4}R_{1,m} = x^{|m|} e^{-\frac{x^2}{4}} e^{-\frac{a}{M}x} H_1(x). \tag{41}$$

The next solution is for, $a_0 \neq 0$, $a_1 \neq 0$, and $a_2 = 0$. This gives

$$a_2 = -\frac{(a - \gamma(M+2))}{2(M+1)} a_1 = 0, \tag{42}$$

$$\gamma_{2,m} = \frac{a}{(M+2)}. \tag{43}$$

The corresponding energies are then,

$$_1\mathcal{E}_{2,m} = \beta(2m+1) - \frac{\alpha^2}{(2m+3)^2}, \tag{44}$$

$$_4\mathcal{E}_{2,m} = \beta - \frac{\alpha^2}{(2|m|+3)^2}. \tag{45}$$

The associated $H$ function is then,

$$H_2 = a_0\left(1 - \frac{(a-\gamma M)}{M}x - \frac{a}{3(2+M)M}x^3 + \cdots\right). \tag{46}$$

Where $\gamma$ is given by Eq. (43) and $a_4$ and beyond are given by Eq. (35). The radial wave function is then

$$_{1\text{ or }4}R_{1,m} = x^{|m|}e^{-\frac{x^2}{4}}e^{-\frac{a}{M+2}x}H_2(x). \tag{47}$$

The next solution is, $a_0 \neq 0$, $a_1 \neq 0$, $a_2 \neq 0$ and $a_3 = 0$,

$$a_3 = -\frac{\bigl(a-\gamma(M+4)\bigr)\bigl(a-\gamma(M+2)\bigr)(a-\gamma M) + 2(M+1)a}{3(M+2)2(M+1)M}a_0 = 0 \tag{48}$$

The energy (recall $\gamma^2 = |\mathcal{E} - m - |m| - 1|$) comes from a solution of the third order polynomial by setting $a_3$ to zero,

$$\bigl(a-\gamma(M+4)\bigr)\bigl(a-\gamma(M+2)\bigr)(a-\gamma M) + 2(M+1)a = 0. \tag{49}$$

The above equation can be solved, for $\gamma_{3,m}$, using the formulas for the roots of a third order polynomial but the real root is not terribly illuminating (the MAPLE output takes up most of a page). It should be noted that in solving the third order polynomial there is one real root and two complex roots. The two complex roots are taken to be unphysical. The radial wave function would then be,

$$H_3 = a_0\left(1 - \frac{(a-\gamma M)}{M}x + \frac{\bigl(a-\gamma(M+2)\bigr)(a-\gamma M)}{2(M+1)M}x^2 + a_4x^4 + \cdots\right). \tag{50}$$

Where, $\gamma$ is given by the real root of Eq. (49), $a_4 = \frac{2(a-\gamma)(a-\gamma M)}{4(M+3)2(M+1)M}$, and $a_5$ and beyond are given by Eq. (35). The radial wave function is then,

$$_{1 \text{ or } 4}R_{3,m} = x^{|m|} e^{-\frac{x^2}{4}} e^{-\gamma_{3,m} x} H_3(x). \tag{51}$$

Notice that the equations for $\gamma_{1,m}$ and $\gamma_{2,m}$ are very simple. $\gamma_{3,m}$, however, is the root of a third order polynomial whose coefficients depend on $a$ and $M$. The energy eigenvalue for this level and beyond will no longer be a simple arithmetic sum of electric and magnetic energies. In general, $_{1 \text{ or } 4}\varepsilon_{k,m}$ is contained in $\gamma_{k,m}$ which is the root of a polynomial of order $n+1$. Once the order of the polynomial is beyond four the root must be found numerically (unless by some quirk the given values for the magnetic field, electric field, and $m$ allow the polynomial to be factored), hence, only the first four states can be guaranteed to be completely determined analytically.

Notice that orthogonal polynomials are not obtained, as is the case for the magnetic field only or the electric field only configurations, but rather, a set of functions of the form

$$e^{-\gamma_{k,m} x} H_k(x), \tag{52}$$

which are orthogonal with respect to the weighting function $x^{|m|} e^{-\frac{x^2}{4}}$ on the interval $[0, \infty)$ for all $k \geq 1$ (we know this because we can write Eqs. (22) and (23) as Sturm-Liouville problems). Also note that the set of orthogonal functions generated by $m > 0$ would be a different set than the set of orthogonal functions generated by $m < 0$.

Cases (2) and (5): $\varepsilon - (m + |m| + 1) = 0$

For these cases the radial equation, Eq. (22) or Eq. (23), has fewer terms but there is no additional asymptotic behavior to remove. Notice also that the energy eigenvalue is known before the radial equation is actually solved

$$_2\varepsilon_m = \beta(2m + 1). \tag{53}$$

This is the same energy for the magnetic field only case with the radial quantum number being zero. For case (5), the magnetic field being repulsive, the energy reduces to

$$_5\varepsilon_m = \beta. \tag{54}$$

It is interesting that the energy eigenvalue has no dependence on $m$ but is dependent only on the strength of the magnetic field.

The radial equation for these two cases is,

$$G'' + \left(\frac{M}{x} - x\right) G' + \frac{a}{x} G = 0. \tag{55}$$

Again, a power series is used to solve the equation

$$G = \sum_{n=0}^{\infty} a_n x^n. \qquad (56)$$

This generates three equations,

$$a_1 = -\frac{a}{M} a_0, \qquad (57)$$

$$a_2 = -\frac{a}{2(M+1)} a_1 = \frac{a^2}{2M(M+1)} a_0, \qquad (58)$$

$$a_n = -\frac{a \cdot a_{n-1} - (n-2)a_{n-2}}{n(M+n-1)}. \qquad (59)$$

Since $a$ cannot be zero this means that $a_1$ cannot be zero unless $a_0$ is zero, and similarly for $a_2$. Now consider $a_3$,

$$a_3 = -\frac{a \cdot a_2 - a_1}{3M+6}, \qquad (60)$$

$$a_3 = -\frac{a(a^2 + 2(M+1))}{3(M+2)2(M+1)M} a_0. \qquad (61)$$

Notice that $a_3$ cannot be zero. Given the alternating sign in the recursion relation it is clear that none of the coefficients can be zero. The radial component of the wave function is then,

$$_{2 \text{ or } 5}R_m = x^{|m|} e^{-\frac{x^2}{4}} \left( 1 - \frac{ax}{M} + \frac{a^2 x^2}{2(M+1)M} - \frac{a(a^2 + 2(M+1))x^3}{3(M+2)2(M+1)M} + \cdots \right). \qquad (62)$$

It is an interesting artifact that the electric and magnetic field are present in the wave function but the electrical field does not appear explicitly in the energy eigenvalue. Notice that even if $M = 1$, i.e. $m = 0$, the series converges. It is also interesting to note that if the electric field is turned off then Eq. (62) becomes,

$$_2 R_m(a=0) = x^{|m|} e^{-\frac{x^2}{4}}. \qquad (63)$$

Which is the Landau state for the radial quantum number being zero (cf. page 252 of Ref. 14). One must wonder if Eq. (62), for $a \neq 0$, is a physical solution or an extraneous solution as turning the electric field on or off has no effect on the energy eigenvalue but does affect the radial wave function.

Cases (3) and (6): $\mathcal{E} - (m + |m| + 1) > 0$

The solutions found by Refs. 1 and 2 are in these two sets. Here the radial equation takes the form (with no additional asymptotic behavior to remove)

$$G'' + \left(\frac{(2|m|+1)}{x} - x\right)G' + \left(\varepsilon - (m+|m|+1) + \frac{a}{x}\right)G = 0. \tag{64}$$

Let $\gamma = \varepsilon - (m+|m|+1) > 0$ and $M = 2|m|+1$, then, for both case (3) and (6) the radial differential takes the following form,

$$G'' + \left(\frac{M}{x} - x\right)G' + \left(\gamma + \frac{a}{x}\right)G = 0. \tag{65}$$

Again, it is assumed that $G$ can be represented by a power series,

$$G = \sum_{n=0}^{\infty} a_n x^n. \tag{66}$$

This generates three equations,

$$a_1 = -\frac{a}{M}a_0, \tag{67}$$

$$a_2 = \frac{\gamma M - a^2}{2a(M+1)}a_1 = \frac{a^2 - M\gamma}{2(M+1)M}a_0, \tag{68}$$

$$a_n = -\frac{a}{n(M+n-1)}a_{n-1} - \frac{(\gamma-n+2)}{n(M+n-1)}a_{n-2}. \tag{69}$$

Notice that this set of recursion relations, Eqs. (67)-(69), has the same issue that the recursion relations, Eqs. (33)-(35), have, setting one of the coefficients of the desired power series to zero yields the eigenvalue and the resulting eigenfunction is not a finite order polynomial but an infinite series. These eigenfunctions will be denoted by $G_k$ with the subscript indicating which coefficient of the power series was set to zero (thus identifying the order of the eigenvalue as well).

For nontrivial solutions, $a_0 \neq 0$ and $a_1 \neq 0$. The first term in the recursion relation that can be zero is $a_2$ (Eq. (68), i.e. there is no $_{3\ or\ 6}R_{0,m}$ or $_{3\ or\ 6}R_{1,m}$),

$$a^2 - \gamma M = 0, \tag{70}$$

$$\gamma_{2,m} = \frac{a^2}{M}, \tag{71}$$

$$_3\varepsilon_{2,m} = \beta(2m+1) + \frac{\alpha^2}{2m+1}. \tag{72}$$

The corresponding energy for the repulsive magnetic field case is then,

$$_6\varepsilon_{2,m} = \beta + \frac{\alpha^2}{2|m|+1}. \tag{73}$$

$G_2$ is then given by

$$G_2 = \left(1 - \frac{a}{M}x + \frac{(\gamma-1)}{3(M+2)}\frac{a}{M}x^3 + \cdots\right)a_0. \tag{74}$$

Where $\gamma$ is given by Eq. (71) and $a_4$ and beyond are given by Eq. (69). The radial solution for both cases (with appropriate $\gamma$ and $M$) is then,

$$_{3 \text{ or } 6}R_{2,m} = x^{|m|}e^{-\frac{x^2}{4}}G_2(x). \tag{75}$$

The next solution can be found by setting $a_3 = 0$. Note that $a_3$ is given by

$$a_3 = \left(\frac{M\gamma - a^2}{3(M+2)2(M+1)} + \frac{(\gamma-1)}{3(M+2)}\right)\frac{a}{M}a_0. \tag{76}$$

The resulting equations for $\gamma$ and the energy eigenvalue are then,

$$\gamma_{3,m} = \frac{2(M+1)}{(3M+2)} + \frac{a^2}{(3M+2)}, \tag{77}$$

$$_3\varepsilon_{3,m} = \beta\left(2m+1+\frac{4(m+1)}{6m+5}\right) + \frac{\alpha^2}{6m+5}. \tag{78}$$

The corresponding energy for the repulsive magnetic field case is then,

$$_6\varepsilon_{3,m} = \frac{(10|m|+9)}{(6|m|+5)}\beta + \frac{\alpha^2}{(6|m|+5)}. \tag{79}$$

$$G_3 = \left(1 - \frac{a}{M}x + \frac{a^2-M\gamma}{2(M+1)M}x^2 + a_4 x^4 + \cdots\right)a_0. \tag{80}$$

Where $\gamma$ is given by Eq. (77) and $a_4$ and beyond are given by Eq. (69). The associated radial solution for both cases can be written down as

$$_{3 \text{ or } 6}R_{3,m} = x^{|m|}e^{-\frac{x^2}{4}}G_3(x). \tag{81}$$

The next order radial solution is found in a similar fashion. Notice that to determine the coefficients of the power series (the $G_k$ functions) in the radial wave function and the energy eigenvalue a polynomial must first be solved for $\gamma$. To be clear, if $_{3\ or\ 6}R_{n,m}$ is desired to be found the polynomial arising from setting the coefficient $a_{n+1}$ equal to zero must be solved for $\gamma$. Due to the structure of the recursion relation, Eq. (69), the order of the polynomial determining $\gamma$ increases by one every other step in the act of solving $a_n = 0$ for $\gamma$. Hence, the last polynomial that can be solved analytically is the one arising from $a_9 = 0$. This generates a fourth order polynomial. The next step, $a_{10} = 0$, generates a fifth order polynomial at which point numerical methods must be used. This means $G_2$ through $G_9$ can be found analytically but the energy eigenvalue associated with $G_{10}$ (and higher order states) and consequently its power series coefficients, must be determined numerically.

For some level of completeness, the energy eigenvalue for $a_4 = 0$ is given below (case 3 only)

$$_3\mathcal{E}_{3,m} = \frac{\alpha^2(3M+4)}{3M(M+2)} + \beta(M+1)$$
$$+ \frac{\beta\sqrt{2\alpha^4(3M^2+9M+8) - \alpha^2 6M(M+2)(M+3) + 9M^2(M+2)^2}}{3M(M+2)}, \quad (82)$$

The complexity of the energy eigenvalue seems to increase rapidly with the order of the wave function. Solutions for these two cases are all of the form ($1 \leq k < \infty$),

$$_{3\ or\ 6}R_{k,m} = x^{|m|}e^{-\frac{x^2}{4}}G_k(x). \quad (83)$$

## V. CONCLUSION

In this paper solutions were found for the two-dimensional hydrogen-like atom in a constant magnetic field turned on adiabatically and perpendicular to the plane of motion for the electron. It was found that the Schrödinger equation, with arbitrary $B$ and $Z$ values, can be solved if the problem is partitioned into six different cases depending on the signs of a shifted energy eigenvalue and the azimuthal quantum number (as this causes the magnetic interaction to be attractive or repulsive). The following radial wave functions were found ($n$ is the radial and $m$ is the azimuthal quantum numbers)

$$_{1\ or\ 4}R_{k,m} = x^{|m|}e^{-\frac{x^2}{4}}e^{-\gamma_{k,m}x}H_k(x), \quad 1 \leq k < \infty, \quad \mathcal{E} < m + |m| + 1, \quad (84)$$

$$_{2\ or\ 5}R_m = x^{|m|}e^{-\frac{x^2}{4}}\sum_{n=0}^{\infty} a_n x^n, \quad a_n \neq 0 \quad \mathcal{E} = m + |m| + 1, \quad (85)$$

$$_{3\text{ or }6}R_{k,m} = x^{|m|}e^{-\frac{x^2}{4}}G_k(x), \quad 2 \leq k < \infty, \quad \varepsilon > m + |m| + 1, \quad (86)$$

where, $H_k$ is given by Eqs. (33)-(35) in $_{1\text{ or }4}R_{n,m}$, $a_n$ is given by Eqs. (57)-(59) in $_{2\text{ or }5}R_m$, and $G_k$ is given by Eqs. (67)-(69) in $_{3\text{ or }6}R_{k,m}$.

An interesting observation here is concerning our ability (or inability) to solve polynomial equations. According the Abel-Ruffini theorem (cf. page 264 of Ref. 18) there is no solution to general polynomial equations of degree five or higher in terms of radicals. In cases 1, 3, 4, and 6 the coefficients of the power series within the eigenfunctions are determined by knowing the energy eigenvalue. The energy eigenvalues are, in effect, roots of polynomials. In cases 1 and 4 this limits us to only the first four energy levels and in cases 3 and 6 to the first eight energy levels for which the energy eigenvalues can be computed exactly for arbitrary magnetic field strengths and $Z$ values. For higher order energy levels, the numerical value of the magnetic field strength, $Z$, and the specific $m$ value must be given before the appropriate polynomial can be solved for the energy eigenvalue and then the subsequent coefficients for the eigenfunction. Cases 2 and 5 are somewhat different in that the energy eigenvalue is known exactly before the eigenfunction is computed. The recursion relations, Eqs. (57)-(59), can determine the coefficients to any desired order in the power series defining the eigenfunctions for these two cases.

The solutions found in this paper answer the question posed by Robnik and Romanovski[8], "*We are looking for solutions at discrete values of the eigen energies that satisfy the Schrödinger integrability condition*

$$\int_0^\infty x^{2|m|}e^{-\frac{x^2}{2}}w^2 x\, dx < \infty$$

*is not known how to determine such solutions and we raise the following question: Will w be a polynomial or some other function?*" (the $w$ of Ref. 8 is the part of $_aR_{k,m}$ written after $x^{|m|}e^{-\frac{x^2}{4}}$ in Eqs. (84)-(86) of this paper).

Finally, it should be noted that the set of functions arising from Eq. (22) and Eq. (23) form different sets of orthogonal functions. The differential equations, Eq. (22) and Eq. (23), each represent a different physical system. One where the magnetic field is attractive and one where the magnetic field is effectively repulsive. Further properties of these solutions will be explored in a later paper.